\documentclass[10pt,a4paper]{iopart}
\usepackage{iopams}  
\usepackage{graphicx,psfrag,bbm,latexsym,color,dcolumn,bm,dsfont,bbm,color,
mathrsfs,bbold,latexsym,amsfonts,amssymb}

\def\defi{{\buildrel \;def\; \over =}}
\newcommand{\be}{\begin{equation}}
\newcommand{\ee}{\end{equation}}

\newcommand{\media}[1]{\langle #1 \rangle}

\begin{document}

\title[Cayley Trees and Bethe Lattices]
{Cayley Trees and Bethe Lattices,\\
a concise analysis for mathematicians and physicists}
\author{M. Ostilli $^{1,2}$}
\address{$^{1}$
Dept. of Computational \& Theoretical Sciences, IIUM,
Kuantan, Pahang, Malaysia}
\address{$^{2}$Statistical Mechanics and Complexity Center (SMC), 
INFM-CNR SMC, Rome, Italy}

\begin{abstract}
We review critically the concepts and the applications of
Cayley Trees and Bethe Lattices in statistical mechanics
in a tentative effort to remove widespread misuse of these simple, 
but yet important - and different - ideal graphs. We illustrate, in particular, 
two rigorous techniques to deal with Bethe Lattices, based respectively on self-similarity
and on the Kolmogorov consistency theorem, linking the latter with the
Cavity and Belief Propagation methods, more known to the physics community.
\end{abstract}

\ead{massimo.ostilli@roma1.infn.it}
\maketitle

\section{Introduction} \label{intro} 
After many years since their introduction, Cayley Trees (CT) \cite{Cayley} 
and Bethe Lattices (BL) \cite{Bethe} still play an important role 
as prototypes of graphs. In fact, even if one can say, 
from the point of graph theory, that 
these ideal graphs
\footnote{
We will see soon that, both CT's and BL's have little numerical availability: 
on one hand a CT is very sensitive to the boundary conditions 
while, on the other hand, a BL, due to the fact that is an infinite graph, cannot be simulated (represented) on a PC.} 
are by now obsolete objects which have been
replaced by more realistic random graphs, like the classical random graph since a few decades
\cite{Classical}~
\footnote{Note that nowadays the name Bethe Lattice is often used to indicate the regular random graph,
which is a tree-like graph, but not an exact tree 
(a tree-like graph is a graph containing only long cycles so that locally it looks like a tree \cite{DMAB}).
However, in this paper we will reserve the name Bethe Lattice to indicate an exact (and then infinite) tree.},
and by complex networks more recently \cite{DMAB}, 
their key feature, that is the fact they are exact tree, \textit{i.e.},
cycles-free, makes CT and BL very instructive examples where exact
calculations can be done \cite{Baxter}.
However, even though there are several excellent works on their applications,
there is still a quite widespread confusion about their exact definition
and use. In particular, by a rapid survey (September 2011), ranging from
Wikipedia to famous textbooks in statistical mechanics, as well as many papers
in referred journals of mathematics or physics, one finds
statements claiming, for example, that the BL is the thermodynamic limit
of the CT, or one finds that a BL is the interior of a large CT, that can
be then analyzed by introducing large but finite subtrees, etc... 
Such false statements and misuses are not just formal mistakes, but serious conceptual
misunderstandings that may lead in turn to fatal errors.
The difference between a CT and a BL was emerged long ago in \cite{Eggarter,Matsuda,Muller,Wu,Turban} 
but, nevertheless, confusion on the subject remained
around over the years, both among mathematicians and physicists. 
We think that the main reason for that is due to an ill mathematical approach to the BL,
and to a scarce communication between the physics and mathematics communities.
Moreover, at the time of the Refs. \cite{Eggarter,Matsuda,Muller,Wu,Turban},
a proper nomenclature for the two kind of graphs was not yet consolidated,
causing further confusion.

The aim of this paper is to give a concise definition of CT and BL
and to illustrate ambiguities-free mathematical tools 
to be used for statistical mechanical models built over CT and BL.
We will show two rigorous techniques to deal with
BL: self-similarity and the Kolmogorov consistency theorem, pointing out that the latter 
is equivalent to the 
Cavity and Belief Propagation methods, more known among physicists.

\section{Cayley Trees and Bethe Lattices: definition and basic properties}
Both CT and BL are simple connected undirected graphs $G=(V,E)$ 
($V$ set of vertices, $E$ set of edges)
with no cycles (a cycle is a closed path of different edges),
\textit{i.e.}, they are trees.  
 
A CT of order $q$ with $n$ shells is defined in the following way.
Given a root vertex $({0})$, we link $({0})$ with $q$ new vertices by means of $q$ edges.
This first set of $q$ vertices constitutes the shell $n=1$ of the CT.
Then, to build the shell $l\geq 2$, each vertex of the shell $l-1$ is linked 
to $q-1$ new vertices.
Note that the vertices in the last shell $n$ have degree $q_n=1$, while
all the other vertices have degree $q_l=q$, $l=0,\ldots,n-1$ (the shell $l=0$ is represented by the single root
vertex ({0})).

The BL of degree $q$ is instead defined as a tree in which any vertex has degree $q$,
so that there is no boundary and no central vertex and, as a consequence,
the main difference between CT and BL is simply that CT is finite while BL is infinite:
\begin{eqnarray}
\label{CTBL}
&&\mathrm{CT}: \quad |V|,|E|<\infty, \\
\label{CTBLb}
&&\mathrm{BL}: \quad |V|,|E|=\infty.
\end{eqnarray}
Note that Eqs. (\ref{CTBL}) and (\ref{CTBLb}) imply an important difference for
the average connectivity $c$ (the connectivity of a given vertex is defined as the
number of edges emanating from it) between a CT and a BL.
In fact, for any finite tree, and in particular a CT,
it holds $|V|=|E|+1$. Therefore 
\begin{eqnarray}
\label{CTBL1}
&&\mathrm{CT}: \quad c=2-\frac{2}{|V|}, \\
\label{CTBL1b}
&&\mathrm{BL}: \quad c=q,
\end{eqnarray}
where in deriving $c$ for the CT case we have used $c=2|E|/|V|$ (valid for any graph). 
In Secs. III and IV we will see that the difference between Eqs. (\ref{CTBL1}) and (\ref{CTBL1b})
has a dramatic consequence in statistical mechanics.
 
In probability theory, an important distinction is in order between
a sequence of probability spaces of increasing size that eventually diverges, and a probability 
space that is infinite by definition.
Similarly, in statistical mechanics, one can be more interested in studying the thermodynamic
limit of the density free energy of the system, or else in studying the physical properties
of a system which is defined from the very beginning as an infinite space (physical or abstract).
Despite only the former kind of infinity leads to a constructive theory of statistical mechanics
and seems physically relevant (in the real world nothing is really infinite, neither the
universe), the second kind of infinity may still be, not only mathematically convenient, but even physically
important (we will see this soon). 
From Eqs. (\ref{CTBL}) and (\ref{CTBLb}), we see that, when we study a model of statistical mechanics built on
a CT, we can have access only to the thermodynamic limit of the system,
while when we study a model on a BL, by definition, we have access only
to the physical properties of the model as defined over an infinite space. 
Even tough most of the models in physics show equivalence between the two different
``kind of infinity'', in the case of CT and BL such equivalence is lost.
The reason for such a difference is easily seen even without entering into
the details of a specific model.
In fact, when we study a model on the CT we need to specify the boundary conditions
of the model, while on the BL, by definition, there are no boundary conditions.
Given a CT of degree $q>2$ with $n$ shells, the number of vertices
on the $l$-th shell is given by $N_l=q(q-1)^{l-1}$, while the total number of vertices of the CT
is $N=q((q-1)^n-1)/(q-2)$. We see therefore that the ratio $N_n/N$ of the CT, in the thermodynamic limit, 
$N\to\infty$,
does not approach zero (for $q>2$). This situation is very different from what
happens in a $d$-dimensional regular lattice box, where the ratio of the 
number of boundary vertices (vertices on the $d-1$ dimensional surface), 
with respect to the total number of vertices, for $N\to \infty$ reaches zero as fast as $1/N^{1/d}$.
Therefore, while the thermodynamic limit of a model built on increasing boxes subsets of $Z^d$, 
is equivalent to the physical properties of the model defined on the infinite space $Z^d$,
the boundary conditions becoming negligible, the thermodynamic limit of a model built on increasing
CT's is not equivalent to the physical properties of the model defined on the infinite space BL
where, by definition, there are no boundaries.
No matter how large $N$ is, the model built on the CT will heavily depend on the boundary conditions, a feature
that makes the model on the CT rarely representative of a real world physical system, so that
the model on the BL is often preferred.
In the next Sections the non equivalence between CT and BL 
will be made concrete with the example of the Ising model.

\section{Ising model on Cayley Trees and on General Trees}
Given a CT $(V,E)$ of degree $q>2$ and $n$ shells,
we want to analyze the Ising model built on it,
having Hamiltonian 
\begin{eqnarray}
\label{H}
H\left[\{\sigma_i\}\right]\defi -J\sum_{(i,j)\in E}\sigma_{i}\sigma_{j}-H_0\sum_i \sigma_i,
\end{eqnarray}
and partition function
\begin{eqnarray}
\label{Z}
Z_n\defi\sum_{\{\sigma_i\}}e^{-\beta H\left[\{\sigma_i\}\right]},
\end{eqnarray}
where $\beta=1/T$ is the inverse temperature, $J$ the coupling constant, $H_0$ the external field,
the $\sigma'$s$=\pm 1$s are the spin variables, and $Z_l$ stands for the partition function
of a CT with $l$ shells.
Let us consider free boundary conditions, and, for simplicity, the case $H_0=0$. Since the CT is finite,
we can in particular start to perform the summation of (\ref{Z}) by summing
over the boundary spins, \textit{i.e.}, by summing over the spins 
on the $n$-th shell. By using $\cosh(\beta J \sigma)=\cosh (\beta J)$ we get
the recursive Equation
\begin{eqnarray}
\label{Z1}
Z_n=Z_{n-1}\left[\cosh (\beta J)\right]^{N_n},
\end{eqnarray}
from which, by iterating, we arrive at
\begin{eqnarray}
\label{Z2}
Z_n=\left[\cosh (\beta J)\right]^{N_n+N_{n-1}+\ldots+N_1},
\end{eqnarray}
that is
\begin{eqnarray}
\label{Z3}
Z_n=\left[\cosh (\beta J)\right]^{|V|}.
\end{eqnarray}
From Eq. (\ref{Z3}) we see that the density free energy $f$
\begin{eqnarray}
\label{f}
-\beta f\defi \lim_{|V|\to \infty}\frac{\log(Z_n)}{|V|}=\log\left[\cosh (\beta J)\right]
\end{eqnarray}
is an analytic function of $\beta$ for any $\beta <\infty$, therefore the Ising model
on the CT does not give rise to a spontaneous magnetization.
Eq. (\ref{f}) has actually a more general counterpart, as it holds for any tree, \textit{i.e.}, with no cycles.
This can be easily seen, for example, by using the high temperature expansion of the free energy
\cite{Domb}.
By using the high temperature expansion, for any graph $G=(V;E)$, $f$ can be written as
\begin{eqnarray}
\label{f1}
-\beta f=\lim_{|V|\to\infty}\frac{|E|}{|V|}\log\left[\cosh (\beta J)\right]+\varphi_G,
\end{eqnarray}
where $\varphi_G$ is the non trivial part of the expansion and is defined
as the sum over all the closed non overlapping paths $C$ of $G$ of $\tanh(\beta J)^{l(C)}$, 
where $l(C)$ is the length of the closed path $C$. Now, if $G$ is a tree, there are no
closed paths, therefore $\varphi_G=0$, and from Eq. (\ref{f1}) it follows
\begin{eqnarray}
\label{f2}
-\beta f=\log\left[\cosh (\beta J)\right]\lim_{|V|\to \infty}\frac{|E|}{|V|}.
\end{eqnarray}
The last factor ${|E|}/{|V|}$ is nothing else than $c/2$,
$c$ being the average connectivity of $G$. By using Eq. (\ref{CTBL1})
we see that the free energy of any tree does not depend on the details of the graph.  
In conclusion, the thermodynamic limit of the Ising model built on trees does not have
a spontaneous magnetization. 
However, it can be shown \cite{Matsuda,Muller,Turban} that the free energy density $f$ of 
a CT with $q>2$ is a non analytic function of the external field $h$ for $T\leq T_{\mathrm{BL}}$,
where $T_{\mathrm{BL}}$ will be introduced in the next Section. 

\section{Ising model on the Bethe Lattice}
We now illustrate how to solve an Ising model on the BL
by using two methods which are free of ambiguities.

\subsection{Self-Similarity method}
Let us consider for simplicity a BL lattice $G$ with degree $q=3$ 
with the exception of one single vertex $({0})$ that instead has degree 2 (see Fig. 1).
\begin{figure}
$\includegraphics[width=0.6\columnwidth,clip]{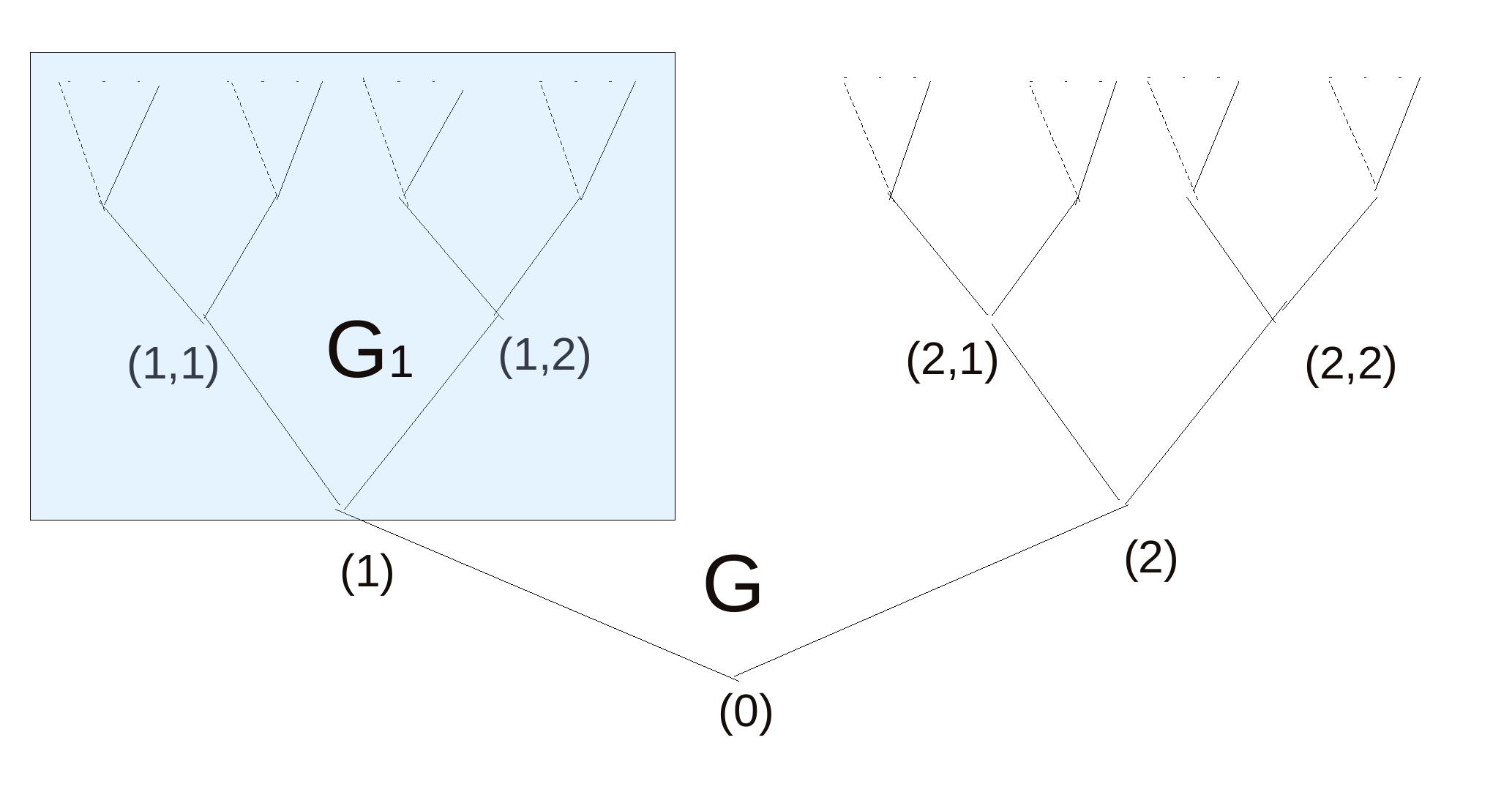}$
\caption{Example of (a portion of) an irregular Bethe Lattice $G$ of degree $q=3$ with the root
vertex ({0}) having only 2 first neighbors. The numbers (1) and (2) in the Figure represent the vertices
in the shell $n=1$, the numbers $(1,1),~(1,2),~(2,1),$ and $(2,2)$, the vertices on the second shell, and so on.
Up to a change of names of the vertices, one has $G=G_1$, where $G_1$ is the (infinite) subgraph in the shaded area.}
\label{f1}
\end{figure}
We introduce this little irregularity on our BL just to make things simpler 
(later on we will show how to restore the fully regular BL).
From the root $({0})$ emanate 2 edges pointing at the vertices $1$ and $2$.
Since $G$ is infinite, it turns out that the two equivalent infinite subgraphs $G_1$ and $G_2$ that we obtain
by eliminating from $G$ the vertex $({0})$ together with its two edges,
are also both equivalent to the original $G$, \textit{i.e.}, up to a change of name of the vertices, 
we have the self-similarity $G=G_1=G_2$. Note, furthermore, that $G_1$ and $G_2$ are each other disconnected. 
If we define the conditional partition function of the system $Z(\sigma_0)$ with respect to the root $({0})$ as 
\begin{eqnarray}
\label{Z2a}
Z(\sigma_0)\defi\sum_{\{\sigma_i, i\neq 0\}}e^{-\beta H\left[\{\sigma_i\}\right]},
\end{eqnarray}
where $H$ is defined similarly to Eq. (\ref{H}), by using the self-similarity and 
the fact that $G_1$ and $G_2$ are disconnected, we get (again for simplicity we consider here
only the case with no external field $H_0=0$)
\begin{eqnarray}
\label{Z3a}
Z(\sigma_0)=\sum_{\sigma_1,\sigma_2}e^{\beta J \sigma_0(\sigma_1+\sigma_2)}Z(\sigma_1)Z(\sigma_2).
\end{eqnarray}
One can feel unease with Eqs. (\ref{Z2a}) and (\ref{Z3a}) due to the fact that G
is infinite, as well as $H$, and the $Z$'s. However we can get rid of the $Z$'s
and any ill defined quantity 
when we consider the finite ratio $P(-)/P(+)=Z(-)/Z(+)$,
where $P(\sigma_0)$ stands for the probability that
the spin at $({0})$ has value $\sigma_0$.
If we define
\begin{eqnarray}
\label{Z4a}
e^{-2h}\defi \frac{P(-)}{P(+)},
\end{eqnarray} 
from Eqs. (\ref{Z3a}) we arrive at the equation for $h$
\begin{eqnarray}
\label{Z5a}
e^{-2h}=\frac{e^{2\beta J-4h}+e^{-2\beta J}+2e^{-2h}}{e^{2\beta J}+e^{-2\beta J-4h}+2e^{-2h}}.
\end{eqnarray} 
Eq. (\ref{Z5a}) has always the trivial solution $h=0$ corresponding to a
zero magnetization.
However, by expanding Eq. (\ref{Z5a}) for small $h$, it is easy to see
that a non trivial solution $h\neq 0$ is present when $T<T_{\mathrm{BL}}$,
where $T_{\mathrm{BL}}$ is the critical temperature given as solution of the equation
\begin{eqnarray}
\label{Z6}
(q-1)\tanh(\beta J)=1,
\end{eqnarray} 
which has solution as soon as $q>2$. 
In other words, the value $q=2$ represents the critical number of neighbors
above which in the system there exists a phase transition.
We can now physically understand why, unlike the BL, in the CT we cannot never have 
a spontaneous magnetization. From Eqs. (\ref{CTBL1}) and (\ref{CTBL1b})
we see that, unlike the BL, on average, no matter how
large the degree $q$ of the CT is, the connectivity $c$ of the CT
is always strictly below the value $c=2$, therefore,
in the CT, no matter how large its degree $q$ is, on average
any spin is surrounded by an insufficient number of neighbors
so that the system cannot have a spontaneous magnetization (at zero external field $H$)~
\footnote{More precisely, since in the thermodynamic limit
for the CT it holds $c=2$, from Eq. (\ref{Z6}) applied with $q=c=2$, we see that the Ising model on the CT is 
critical only at $T=0$.}. Similar conclusions apply to the Potts model \cite{Turban}.
The average magnetization of the spin $\sigma_0$ can be calculated from Eq. (\ref{Z4a}) and gives 
$\media{\sigma_0}=\tanh(h)$, while, the average magnetization for the fully regular BL with $q=3$
is simply $\media{\sigma_0}=\media{\sigma}=\tanh(h\times 3/2)$. Notice however the necessity to consider 
the irregular BL: self-similarity, and then the possibility to have
an equation for the effective field $h$, applies only to the irregular BL, not to the 
fully regular BL. The physicist familiar with the Cavity Method (see next subsection) 
recognizes in this step another way to see that $h$ is the strength felt by a spin due to all the other spins 
in the absence (``the cavity'') of the spin itself.

We end this subsection with the following critical note.
In many textbooks of statistical mechanics, as well as in many
papers that deal with a BL (regular or not), in the place of Eqs. (\ref{Z3a}) or (\ref{Z5a}), 
we find respectively
\begin{eqnarray}
\label{Z3b}
Z_{n+1}(\sigma_0)=\sum_{\sigma_1,\sigma_2}e^{\beta J \sigma_0(\sigma_1+\sigma_2)}Z_n(\sigma_1)Z_n(\sigma_2),
\end{eqnarray}
and
\begin{eqnarray}
\label{Z5b}
e^{-2h_{n+1}}=\frac{e^{2\beta J-4h_n}+e^{-2\beta J}+2e^{-2h_n}}{e^{2\beta J}+e^{-2\beta J-4h_n}+2e^{-2h_n}},
\end{eqnarray} 
and it is often wrongly said that
$Z_n$ stands for the partition function of that finite portion of the BL having only $n$ shells.
From what we have seen all above, it should be clear that such an approach
for the BL is conceptually quite wrong: $Z_n$ cannot be interpreted as the partition function
of a finite subgraph of the BL, otherwise for $Z_n$ we would get again Eq. (\ref{Z3}). 
At this level, the only correct meaning we can attribute here to $Z_n(\sigma)$
is that of the conditional partition function of an infinite subgraph self-similar to that
associated to $Z_{n+1}(\sigma)$, 
as showed above, and, most importantly,
$Z_{n+1}(\sigma)$ and $Z_n(\sigma)$ are the same thing.       
In this context, the index $n$ appearing in Eqs. (\ref{Z3b}) or (\ref{Z5b}) is a quite misleading symbol.
More precisely, one should use instead $Z(\sigma_{n+1})$ and $Z(\sigma_n)$ but, as a consequence, 
for the ratios we would get the same value regardless of the index: $h_n=h_{n+1}$.
Rather, looking at Eqs. (\ref{Z3a}) or (\ref{Z5a}), as a mere numerical tool, it can be convenient
to introduce the recurrent Eqs. (\ref{Z3b}) or (\ref{Z5b}) since its fixed points are equivalent
to Eqs. (\ref{Z3a}) or (\ref{Z5a}), respectively. 
A more rigorous and general meaning to $h_n$ can be attributed by following the next measure-theoretical approach.

\subsection{Kolmogorov' s condition}
A more rigorous approach to the BL comes from the Kolmogorov consistency theorem 
(see \textit{e.g.}, \cite{Shi} and \cite{Sinai})
by which we can avoid to introduce any ill defined quantity but, most importantly, as we will see,
the method attributes a rigorous and physical meaning to Eq. (\ref{Z5b}) and to its generalizations. 

From the point of view of probability theory, 
solving the Ising model on the BL amounts to find all the marginal probabilities
of an infinite probability space, $\textit{i.e.}$, a probability space 
characterized by an infinite
number of random variables (in our case the spins with indices in $V$), 
a suitable sigma-algebra \cite{Shi}, and a measure $\mu$. 
The marginal probabilities can be calculated in turn from an infinite set of given 
finite dimensional marginal distributions $\mu_n=\mu_{|V_n}$, where $V_n$ is the set of vertices that
are at distance not greater than $n$ from 0. Note however that, when we consider $\mu_n$, 
we are not dealing with a finite disconnected subgraph of the given BL; the spins
of $V_n$ that are located on the shell $n$ are in fact in turn connected to the
spins of $V_{n+1}$. The key point to be used is that, given the values of the spins on $V\setminus V_n$,
the spins on $V_n$ are conditioned only by the neighboring spins, \textit{i.e.}, 
the spins of the $n+1$-th shell.  
In turn, given a spin $\sigma_i$, with $i\in V_n$, the effects of all its neighbor spins 
that are on $V\setminus V_n$, can be encoded via
a single effective external field $h_i$ acting only on $\sigma_i$.
Therefore, for the $\mu_n$, we look for solutions of the form 
\begin{eqnarray}
\label{K0}
\mu_{n}(\{\sigma_i, i\in V_n\})=e^{-\beta H_n+\sum_{i\in W_n}h_i\sigma_i}/Z_n
\end{eqnarray} 
where $H_n$ is given by Eq. (\ref{H}), $W_n=V_{n}\setminus V_{n-1}$ stands for the set of vertices
of the shell $n$, and $Z_n$ is the normalization constant.
Physically, Eq. (\ref{K0}) represents the first step of the Bethe-Peierls 
approach \cite{Bethe} (nowadays also known as Cavity-Method in physics \cite{MezardP}, or Belief-Propagation in
computer science \cite{Yedida}) in which the effects of all the spins other than those
of the set $V_n$ are encoded in the fields $\{h_i\}$ to be determined self-consistently.
From a strict probabilistic point of view, Eqs. (\ref{K0}) and (\ref{H}) 
represent the definition of the Ising model itself on the BL.
Now, as we have mentioned in Sec. II, the distinction between
a sequence of probability spaces of increasing size and an infinite probability space
is important. The sequence of probability spaces may or may not converge to anything,
and even if it converges, it may happen that it does not converge to a probability space.
The tool to investigate whether a sequence of increasing probability spaces converges
to an infinite probability space is provided by the Kolmogorov consistency theorem \cite{Shi}.
It should be clear that on an infinite space like the BL, the probability that
a given configuration of spins $\{\sigma_i, i\in \mathbb{N}\}$ (a point-like event) is realized is zero:
$\mu(\{\sigma_i, i\in \mathbb{N}\})=0$. Therefore,
when one considers an infinite probability space, the introduction of a proper sigma-algebra 
is not just a formal necessity. However, since here we do not need to work directly with the 
measure $\mu$ of the infinite space, for what follows, we can skip the introduction of the sigma-algebras.
The Kolmogorov consistency theorem says that a sequence of probability spaces $(V_n,\mu_n)$ 
converges to the probability space on BL, $(V,\mu)$, 
if for any $n\geq 1$ one has 
\begin{eqnarray}
\label{K}
\sum_{\{\sigma_i, i\in W_{n+1}\}}\mu_{n+1}(\{\sigma_i,i\in V_{n+1} \})=\mu_{n}(\{\sigma_i,i\in V_{n} \}).
\end{eqnarray} 
Eqs. (\ref{K0}) and (\ref{K}) give the functional equations for the fields $h_i, i\in W_n$.

Due to the tree-like structure of the BL, Eq. (\ref{K}) factorizes over independent branches
allowing for a fundamental simplification.
Let us apply Eqs. (\ref{K0}) and (\ref{K}) to the same model analyzed in the previous subsection 
(a BL with $q=3$ in which the root vertex ({0}) has only two first neighbors) \cite{Georgii,Bleher}.
For any given $i_0\in W_n$ we indicate with $i_1$ and $i_2$  
the indices of the first two neighbors of $i_0$ that belong to $W_{n+1}$.
We have ($H_0=0$)
\begin{eqnarray}
\label{K1}
&&\frac{e^{-\beta H_n(\{\sigma_l,l\in V_n\})}}{Z_{n+1}} 
\prod_{i_0\in W_n}\sum_{\sigma_{i_1},\sigma_{i_2}}
\left[e^{\beta J\sigma_{i_0}(\sigma_{i_1}+\sigma_{i_2})+h_{i_1}\sigma_{i_1}+h_{i_2}\sigma_{i_2}}\right] \nonumber \\  
&=&\frac{e^{-\beta H_n(\{\sigma_l,l\in V_n\})}}{Z_{n}}\prod_{i_0\in W_n}
e^{h_{i_0}\sigma_{i_0}}~~
\end{eqnarray} 
From Eq. (\ref{K1}), for any $i_0\in W_n$ and for any values of $\sigma_{i_0}$, we have an equation for the field $h_{i_0}$
as a function of the fields $h_{i_1}$ and $h_{i_2}$
\begin{eqnarray}
\label{K2}
\nonumber
&&\frac{Z_n}{Z_{n+1}}
\sum_{\sigma_{i_1},\sigma_{i_2}}e^{\beta J\sigma_{i_0}(\sigma_{i_1}+\sigma_{i_2})+h_{i_1}\sigma_{i_1}+h_{i_2}\sigma_{i_2}}
=e^{h_{i_0}\sigma_{i_0}},
\end{eqnarray} 
from which, by evaluating the cases $\sigma_{i_0}=+1$ and $\sigma_{i_0}=-1$ we get
\begin{eqnarray}
\label{K3}
&&e^{-2h_{i_0}}=\\ 
&&\frac{e^{2\beta J-h_{i_1}-h_{i_2}}+e^{-2\beta J+h_{i_1}+h_{i_2}}+e^{h_{i_1}-h_{i_2}}+e^{h_{i_2}-h_{i_1}}}
{e^{2\beta J+h_{i_1}+h_{i_2}}+e^{-2\beta J-h_{i_1}-h_{i_2}}+e^{h_{i_1}-h_{i_2}}+e^{h_{i_2}-h_{i_1}}}. \nonumber
\end{eqnarray} 
It is immediate to verify that Eqs. (\ref{K3}) reduce to Eq. (\ref{Z5b}) for the choice $h_{i_0}=h_{i_1}=h_{i_2}=h_n$.
Besides to be rigorous, this approach, based on the Kolmogorov consistency theorem, 
shows us that a possible
non homogeneous solution for the effective fields $\{h_i\}$, periodic or not, is not
an artifact of the theory: an effective field which depends on the index vertex is in correspondence
with a non homogeneous marginal probability which depends on the index vertex too. 

In the context of the Cavity and Belief Propagation methods, Eqs. (\ref{K3}) are better known
in another form more suited for an intuitive interpretation of the effective fields as messages
$h_{i\to j}$ passing from a vertex $i$ to a neighbor vertex $j$; see Fig. (2). 
\begin{figure}
$\includegraphics[width=0.4\columnwidth,clip]{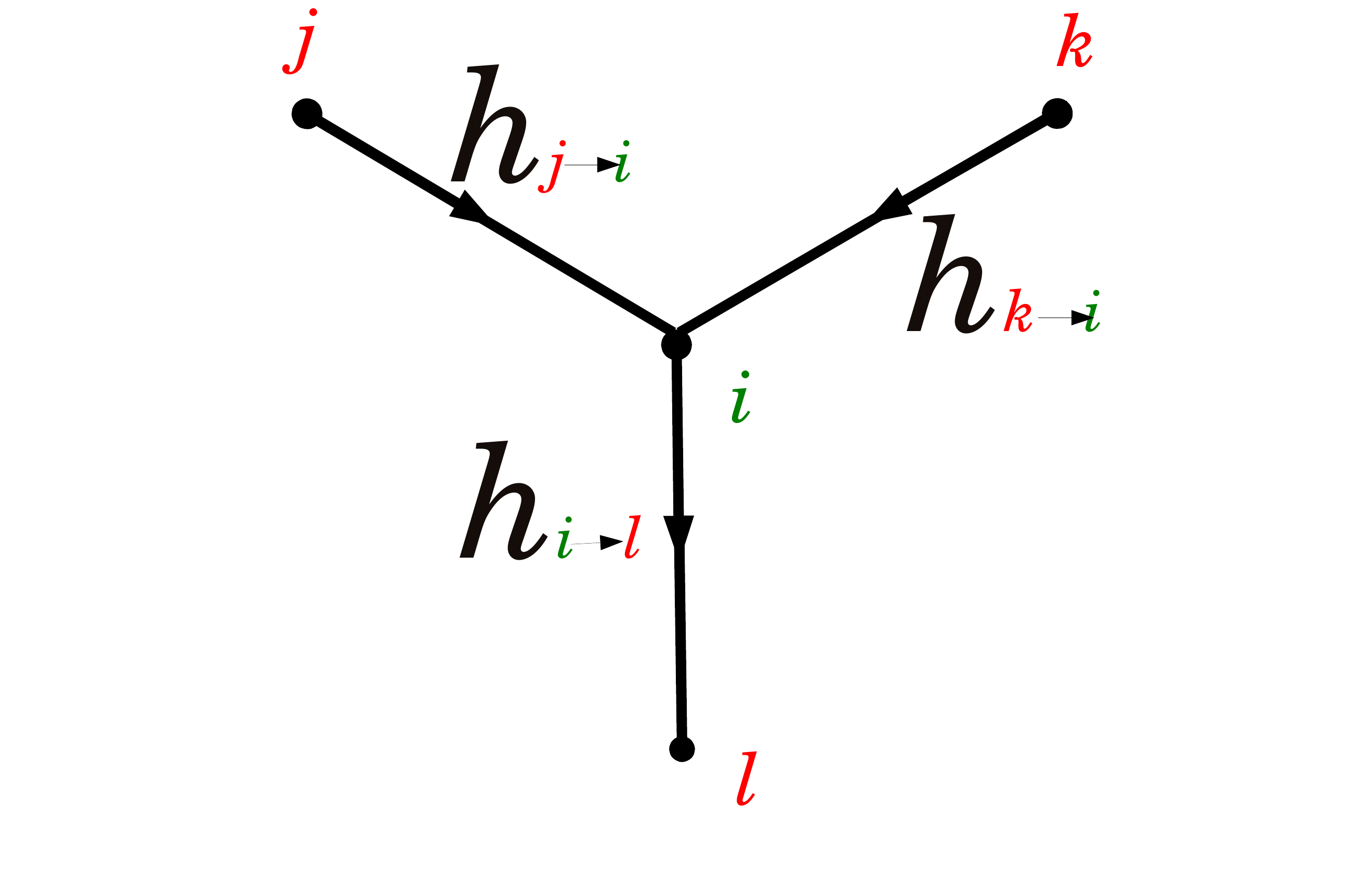}$
\caption{Graphical message-passing interpretation of the effective fields. Each field $h_{i\to l}$
is in correspondence with an edge $(i,l)=(l,i)\in E$ (recall that our BL $G$ is an undirected graph). 
The field/message $h_{i\to l}$, emanating from $i$, as seen from Eq. (\ref{K4}), is a function of all the 
other fields/messages arriving to $i$: $h_{j\to i}$ and $h_{k\to i}$. 
The correspondence between the fields appearing in Eqs. (\ref{K3}) and those of Eqs. (\ref{K4}) is: 
$h_{i\to l}=h_{i_0}$, $h_{j\to i}=h_{i_1}$, and $h_{k\to i}=h_{i_2}$.
Note that, in general, we can attribute a direction to the fields/messages $h_{i\to l}$ even if the 
graph $G$ is an undirected graph (\textit{i.e.}, for any $(i,j)\in E$ we have $(i,j)=(j,i)\in E$).} 
\label{f2}
\end{figure}
By using $\tanh^{-1}(x)=1/2\log[(1+x)/(1-x)]$, valid for $|x|<1$, it is easy to see that
Eqs. (\ref{K3}) can be rewritten as
\begin{eqnarray}\fl
\label{K4}
h_{i\to l}&=&\tanh^{-1}\left[\tanh(\beta J)\tanh(h_{j\to i})\right] 
+\tanh^{-1}\left[\tanh(\beta J)\tanh(h_{k\to i})\right].
\end{eqnarray} 

More in general, the equation for a an Ising model built on a generic BL, regular or not, having generic
couplings $J_{ij}$ and an external field $H_0$, reads as  
\begin{eqnarray}
\label{K4b}
h_{i\to j}=\sum_{k\in \partial i\setminus j} \tanh^{-1}\left[\tanh(\beta J_{ij})\tanh(\beta H_0+h_{k\to i})\right],~~
\end{eqnarray} 
where $\partial i$ stands for the set of the first neighbors of the vertex $i$.

The developments about the convergence and algorithmic issues around Eqs. (\ref{K4b}), 
as well as their applications in physics, computer science, and statistical inference are
huge (see \textit{e.g.} \cite{MezardP}, \cite{Zecchina}, \cite{Yedida}). 
Here we wanted just to stress the equivalence between the Cavity/Belief-Propagation approaches
and a rigorous method based on the Kolmogorov consistency theorem.

\section{Last observation}
From the previous Section, we see that we could re-interpret the (infinite) BL,
as the thermodynamic limit of (finite) CT's having specific ``boundary conditions'' determined by
the fields $h_i$ solution of Eqs. (\ref{K0}) and (\ref{K}), as shown in \cite{Thompson}. 
However, this point of view is a quite poor physical one 
(a system whose thermodynamic limit exists only for a specific boundary condition) 
and, pedagogically, very inconvenient.

\section{conclusions}
We have reviewed critically the concepts and the applications of
CT and BL in statistical mechanics emphasizing their very different features.
We have pointed out that careful must be paid especially in the case of a BL, 
an infinite space, 
where serious misuses and misunderstandings are currently seen in both textbooks 
and journal papers due to a ill mathematical approach to the BL. 
We have then illustrated for the BL case two alternative approaches which
are rigorous and free of dangerous ambiguities based respectively on self-similarity 
and the Kolmogorov consistency theorem, pointing out the link of the latter with
the Cavity and Belief Propagation methods, more known to the physics community.
We hope that this critical review paper, where concepts and tools from 
physics and mathematics (too often kept apart) are used,
might reduce the quite widespread misuse and conceptual errors around CT and BL. 

\section*{Acknowledgments}
Work supported by Grant IIUM EDW B 11-159-0637.
We thank F. Mukhamedov for useful discussions.
%
%

\newpage

\end{document}